\documentclass[final,times,5p,lefttitle]{elsarticle}

\journal{Nuclear Inst. and Methods in Physics Research, A}

\usepackage{docmute}

\begin{document}

\begin{frontmatter}

\title{Imaging and Spectral Performance of a 60 $\mu$m Pitch CdTe Double-Sided Strip Detector}


\author[ut,kavli]{Kento Furukawa}
\author[ut,kavli]{Shunsaku Nagasawa\corref{cor1}}
\author[umn]{Lindsay Glesener}
\author[kavli]{Miho Katsuragawa}
\author[kavli]{Shin'ichiro Takeda}
\author[isas,kavli]{\\Shin Watanabe}
\author[kavli,ut]{Tadayuki Takahashi}

\address[ut]{Department of Physics, The University of Tokyo, 7-3-1 Hongo, Bunkyo, Tokyo 113-0033, Japan}
\address[kavli]{Kavli Institute for the Physics and Mathematics of the Universe(Kavli IPMU, WPI), The University of Tokyo,\\ ~~5-1-5 Kashiwanoha, Kashiwa, Chiba 277-8583, Japan}
\address[umn]{University of Minnesota, Minneapolis, Minnesota, USA}
\address[isas]{Institute of Space and Astronautical Science, Japan Aerospace Exploration Agency (ISAS/JAXA),\\ ~~3-1-1 Yoshinodai, Chuo-ku, Sagamihara, Kanagawa 252-5210, Japan}
\cortext[cor1]{E-mail address: shunsaku.nagasawa@ipmu.jp}




\begin{abstract}
We have evaluated the performance of a fine pitch CdTe Double-sided Strip Detector (CdTe-DSD), which was originally developed for the focal plane detector of a hard X-ray telescope to observe the Sun.
The detector has a thickness of 750 $\mathrm{\mu}$m and has 128 strip electrodes with a 60 $\mathrm{\mu}$m strip pitch orthogonally placed on both sides of the detector and covers an energy range 4 keV to 80 keV.
The study of the depth of photon interaction and charge sharing effects are of importance in order to provide good spectroscopic and imaging performance.
We study the tail structure observed in the spectra caused by charge trapping and develop a new method to reconstruct the spectra based on induced charge information from both anode and cathode strips.
By applying this method, energy resolutions (FWHM) of 0.76~keV and 1.0~keV can be obtained at photon energies of 14~keV and 60~keV,  respectively, if the energy difference between the anode and cathode is within 1~keV.
Furthermore, the tail component at 60~keV is reduced, and the energy resolution of the 60~keV peak is improved from 2.4~keV to 1.5~keV (FWHM) if the energy difference is greater than 1~keV.
In order to study the imaging performance, we constructed a simple imaging system using a 5~mm thick tungsten plate that has a pinhole with a diameter of 100 $\mathrm{\mu}$m.
We utilize a ${}^{133}$Ba radioisotope of 1~mm in diameter as a target source in combination with a 100 $\mathrm{\mu}$m slit made from 0.5 mm thickness tungsten. 
We imaged the ${}^{133}$Ba source behind the 100 $\mathrm{\mu}$m slit using a 30 keV peak, with a 100 $\mathrm{\mu}$m pinhole placed at the center of the source-detector distance.
By applying a charge sharing correction between strips, we have succeeded in obtaining a position resolution better than the strip pitch of 60 $\mathrm{\mu}$m.

\end{abstract}


\end{frontmatter}


\section{Introduction}
Semiconductor detectors  for spectroscopy and imaging of hard X-ray photons at several tens of keV are now playing an important role in various research fields, from X-ray astronomy to non-destructive material analysis and in vivo medical imaging. 
Recently high resolution, photon counting, imaging spectrometers have emerged as attractive technologies for these applications, in which images and spectra are obtained simultaneously from the interaction position and the energy of each photon.

To achieve high energy resolution and high position resolution at the same time, we have been developing CdTe Double-sided Strip Detectors (CdTe-DSD), based on a CdTe diode\cite{takahashi1999high,takahashi2001recent}, ranging from a strip pitch of 250 $\mathrm{\mu}$m to 60 $\mathrm{\mu}$m \cite{Watanabecdte2009,takahashi2017high}, in combination with dedicated readout electronics \cite{takahashi2001high, Nakazawa2004, Takahashi2005}.
Imaging is achieved by orthogonal strip electrodes on the two sides of the detectors\cite{Katsuragawa2018, Takeda2018, Ishikawa2016}. 

One merit of the DSD configuration is that signals can be processed with a smaller number of electronic channels than in the case of a pixel configuration, which generally means more space for complex circuits for low noise amplification. Furthermore, signals generated by the movement of carriers in the device can be measured from both the anode strips and the cathode strips. $\mu\tau$ products in CdTe are known to be small in comparison with other semiconductor devices, such as Si and Ge, and there is a factor of about ten in this product between electrons and holes.
Therefore the total amount of induced charge for a given photon energy depends on the distance between the interaction point and the electrode due to the hole trapping in the device and it is expected that these signals carry information regarding the depth of interaction.

Another piece of information we can obtain from the configuration of fine pitch strips is the charge spread. It has been reported that  the charge clouds generated in sensors are spread across multiple pixels, if pixel sizes are reduced to the extent of the clouds\cite{Koenig2013}.
This suggests that the  spread of the cloud over multiple strips on the DSD could be used to reconstruct the interaction location in the detector which improves the position resolution to a level finer than the strip pitch.\\

Following onto our previous publication\cite{Furukawa2018}, we have performed a detailed study of the characteristics of a fine pitch CdTe-DSD using 60 $\mathrm{\mu}$m pitch strips, which was originally developed for the focal plane detector for a hard X-ray telescope used in the third flight of a sounding rocket experiment, named the Focusing Optics X-ray Solar Imager (FOXSI-3) which was used to observe the Sun\cite{Glesener2016,Musset2019}. 
In this paper, we report significant improvements in the spectral performance by comparing the pulse height measured from anode and cathode strips to correct  the effects of depth of interaction, caused by  hole trapping. We also report the improvement of image resolution by properly taking the charge spread into account.

\section{Detector Configurations}
\begin{figure}[htb]
\begin{center}
\includegraphics[width=1.0\hsize]{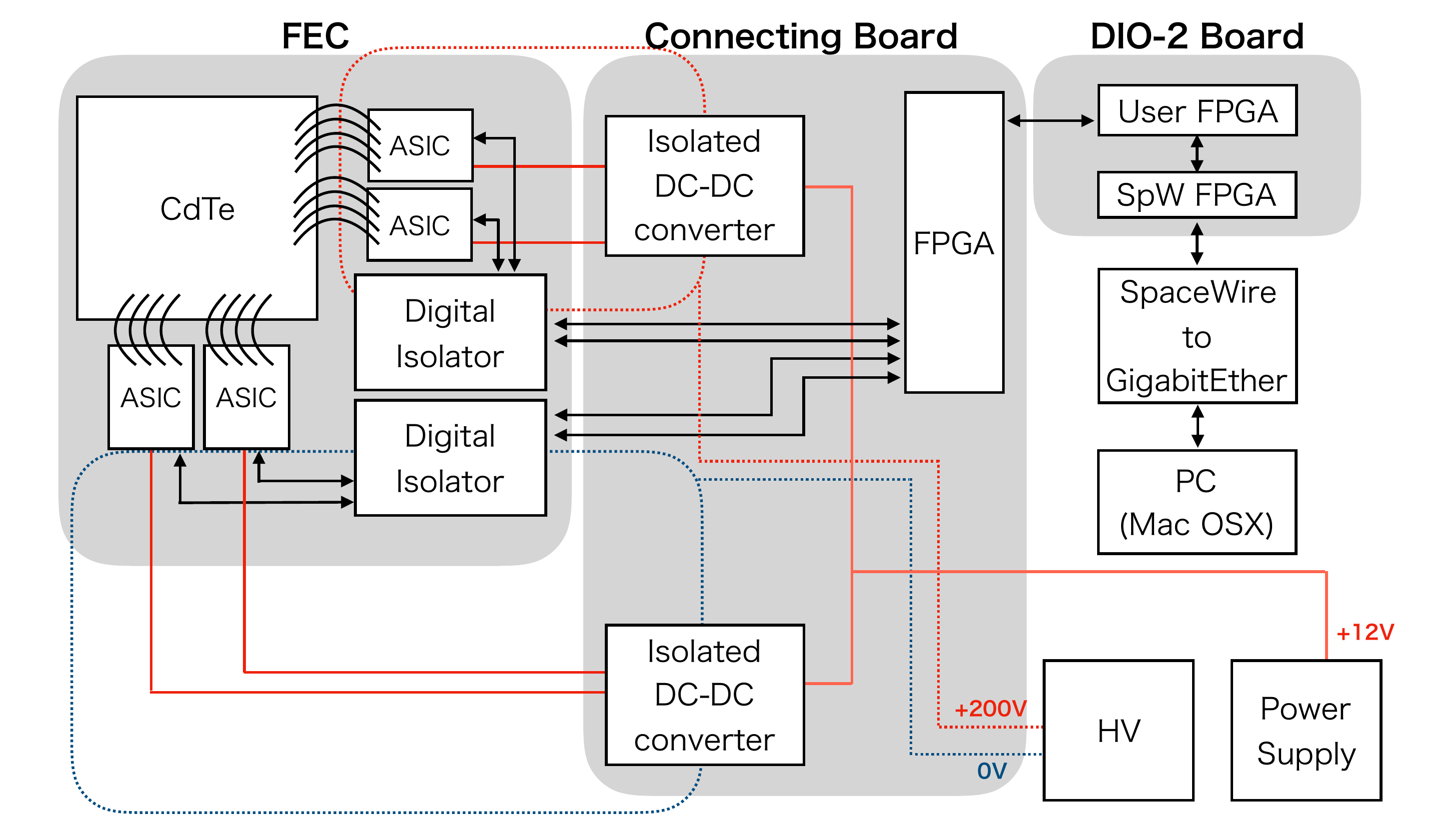}
\includegraphics[width=1.0\hsize]{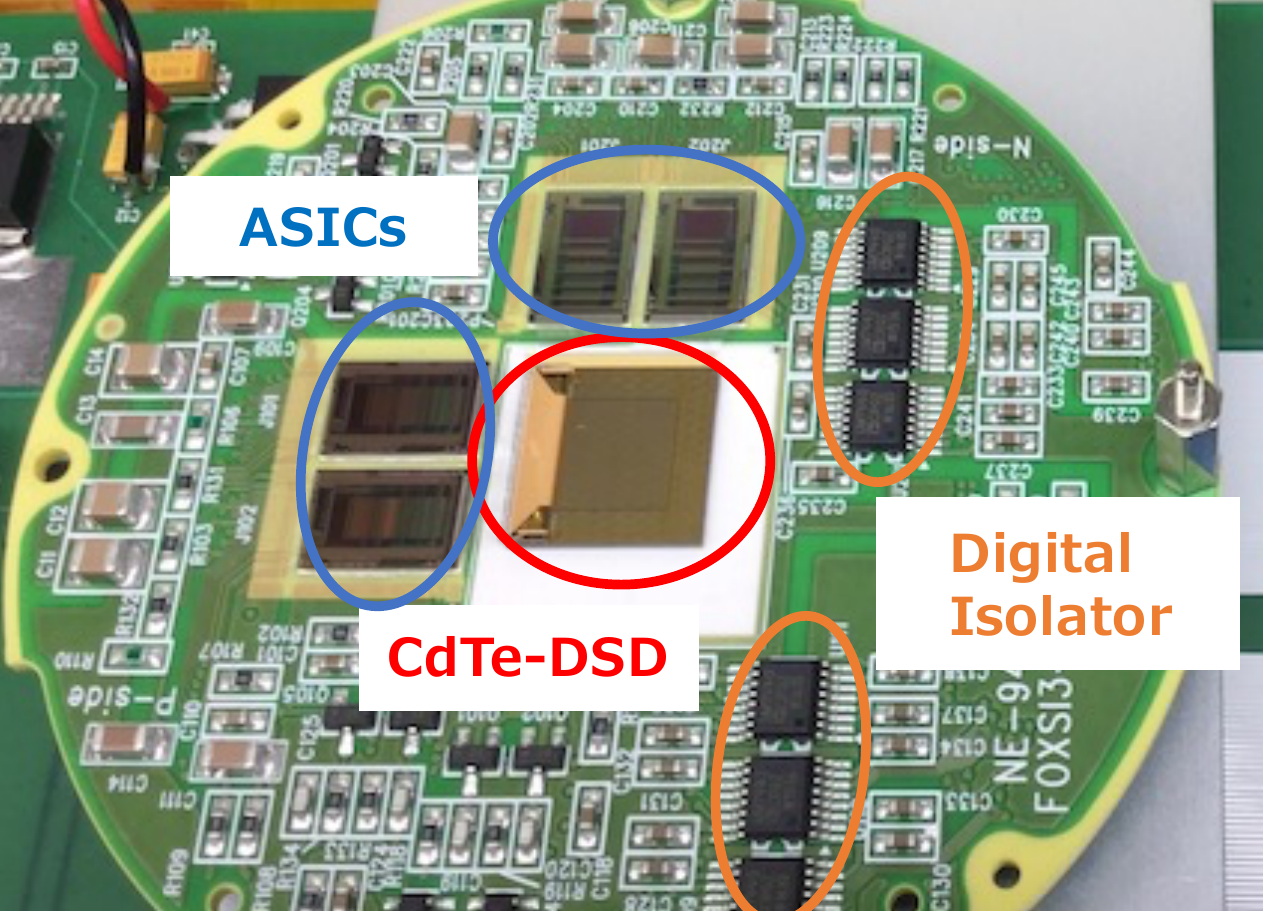}
\caption{A picture of the 60 $\mu$m pitch CdTe-DSD and a system diagram for the calibration data acquisition system. A round Front End Card (FEC) is wired to a Connecting Board. The CdTe device is placed at the center of the FEC, which is connected to four VATA-SGD ASICs originally developed for the Soft Gamma-ray Detector (SGD) onboard the ASTRO-H satellite \cite{Watanabe2014}.}
\label{system}
\end{center}
\end{figure}

Our experiment uses a detector made for the FOXSI-3 experiment.
Figure~\ref{system} shows a picture of a Front End Card (FEC) of the 60 $\mu$m pitch CdTe-DSD for FOXSI and a system diagram of an experimental setup for calibration data acquisition.
The detector has a thickness of 750 $\mathrm{\mu}$m and covers an energy range of 4~keV to 80~keV. 
It has 128 strip electrodes on each side and the width of each strip electrode is 50~$\mu$m with a 10~$\mu$m gap between the strips.
The imaging area is 7.68 $\times$ 7.68 mm${}^{2}$.
More details are described in \cite{Furukawa2018}.

The CdTe sensor is connected to four Application Specific Integrated Circuits (ASICs) for pulse height measurement.
Each ASIC has 64 channels and each channel contains a preamplifier, fast shaper for self-triggering and slow shaper for pulse height measurement\cite{Watanabe2014}.
Anode and cathode electrodes are placed orthogonal to each other and pulse heights from both sides reflect 2-dimensional information on where X-rays intersect the device.

\section{Spectroscopic Performance}
\subsection{Anode and Cathode spectrum of ${}^{241}$Am}
\begin{figure}[htb]
\begin{center}
\includegraphics[width=\hsize]{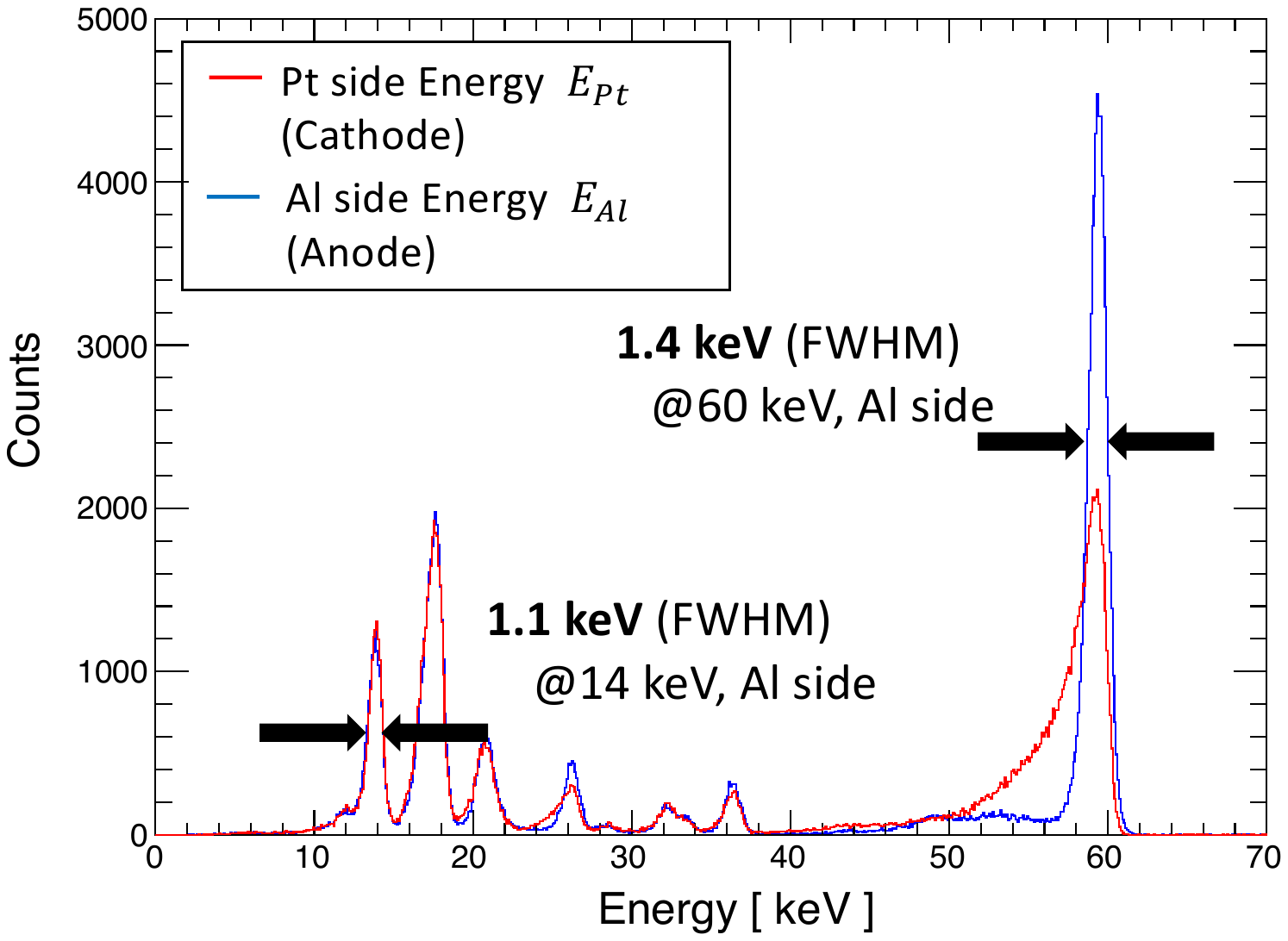}
\caption{Spectra of $^{241}$Am, summed over the detector. Red: Energy spectrum observed on the Pt(cathode) side. Blue: Energy spectrum observed on the Al(anode) side.}
\label{before_spect}
\end{center}
\end{figure}

The CdTe-DSD provides two independent measurements of energy from its anode and cathode electrodes. Figure~\ref{before_spect} shows two $^{241}$Am spectra obtained by the two sides of the detector.
The detector was irradiated on the cathode side.
If energy is detected on two or more adjacent strips (multi-strip event), the total sum of energies on the adjacent strips are used.
In the anode side spectrum, energy resolutions are 1.1~keV at 14~keV and 1.4~keV at 60~keV in FWHM.
Below 20~keV, the cathode side spectrum is almost identical to that of the anode side.
In contrast, the peaks above 20~keV exhibit low-energy tails.
This tail structure is caused by the depth dependence of the induced charge due to the charge trapping effect\cite{takahashi2001recent,salcin2014,He2001}.

\begin{figure}[htb]
\begin{center}
\includegraphics[width=\hsize]{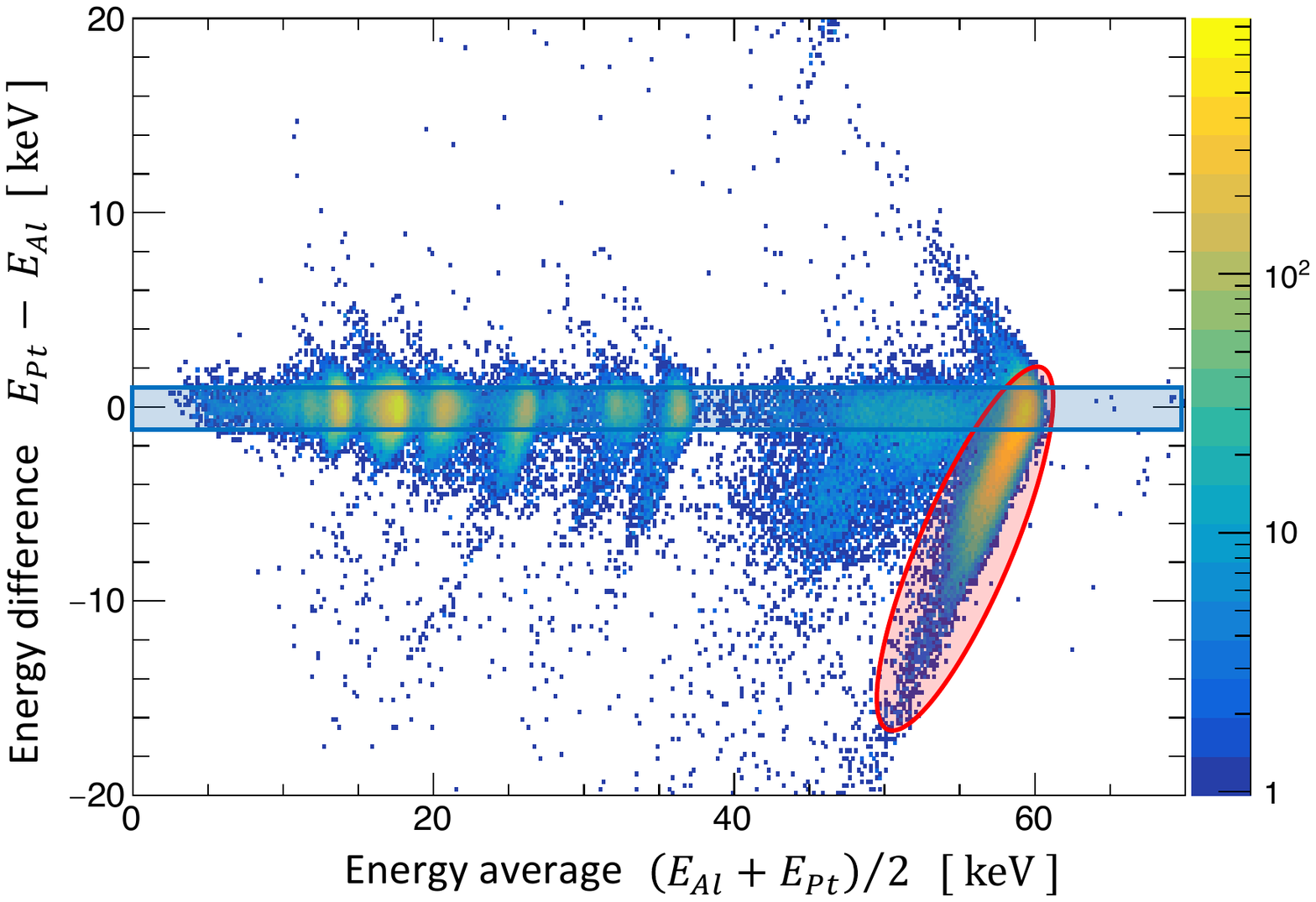}
\caption{The correlation between the anode side energy and the cathode side energy. The x-axis is the average of the anode and cathode side energies. The y-axis is the difference between them. The margin ($\pm$1~keV$,\pm1.5\sigma$) of the combined energy resolution is indicated by the blue rectangle. The tail structure at 60~keV peak is shown in the red region.}
\label{avediff}
\end{center}
\end{figure}

When high energy photons ($>$20~keV) enter the detector, some of the photons penetrate relatively deeply into the detector and the depth of interaction (DoI) is a vertically spread distribution.
As the DoI becomes deeper, the total measured energy on the cathode electrodes gets smaller, creating a tail on the peak toward low energies.
Figure~\ref{avediff} compares the energies measured by the cathode and anode sides. The low-energy tail on the 60~keV cathode-side peak is indicated by the red region.
For low energies ($<$20~keV), the anode and cathode energy spectra are almost identical and averaging both measurements results in the expected energy resolution improvement\cite{Watanabecdte2009}.
However, for higher energies ($>$20~keV), a more complicated correction is required to improve the energy resolution. We have applied an algorithm applicable up to at least 60~keV, confirming the spectroscopic performance improvement across the entire range.

\subsection{Averaged Spectra}
\begin{figure}[htb]
\begin{center}
\includegraphics[width=\hsize]{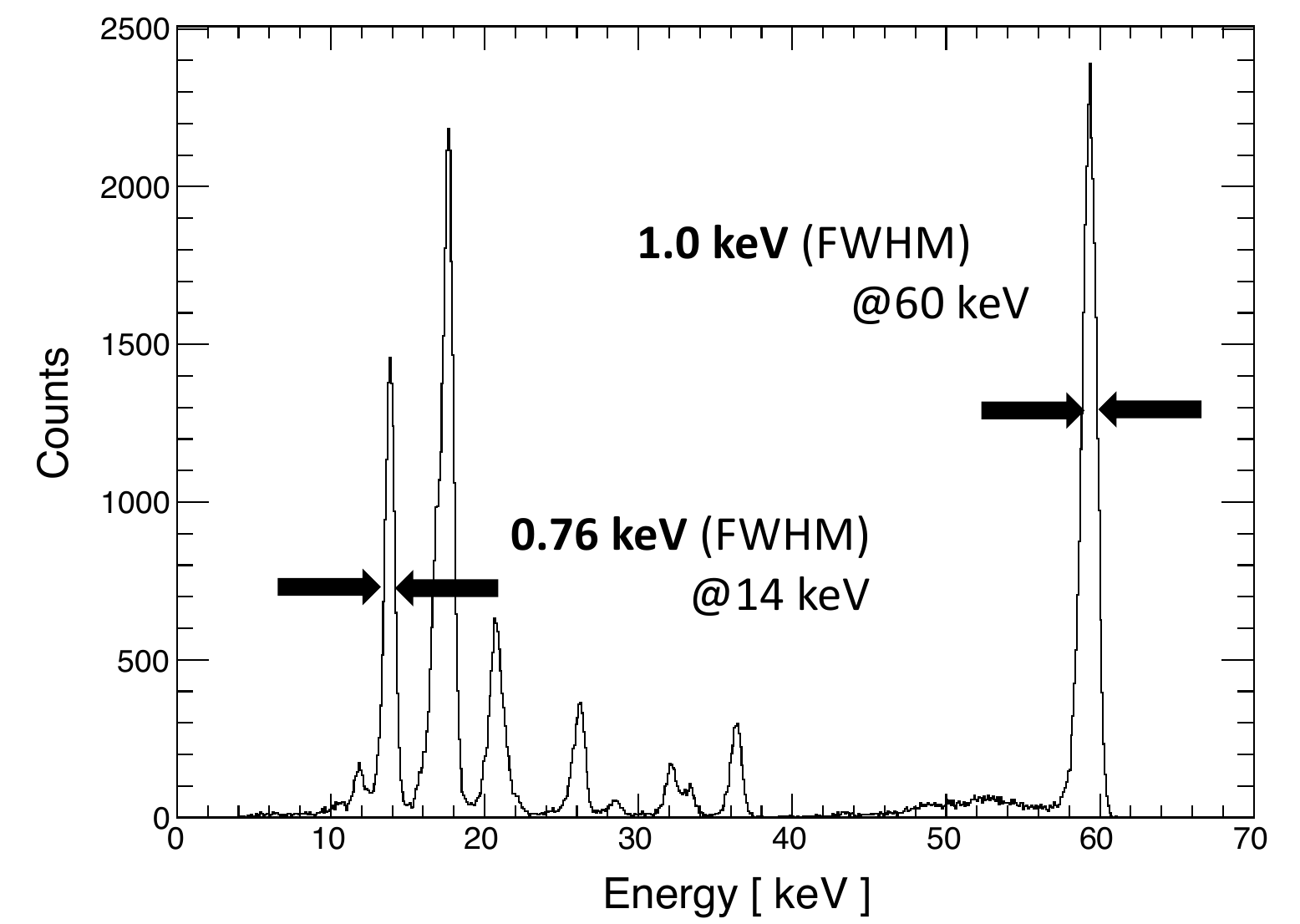}
\caption{Spectrum obtained by averaging the anode and cathode energy when the difference is within the energy resolution. The improvement factor of the energy resolution is consistent with the statistically expected value of $\sqrt{2}$.}
\label{ave_when_diff_small}
\end{center}
\end{figure}

When the anode side energy and the cathode side energy are equal within a certain margin, a simple average works well to obtain a superior energy resolution.
Figure~\ref{ave_when_diff_small} shows the averaged spectrum when the difference between the anode and cathode side energy is within the energy resolution.
The energy resolution at 14~keV is 1.1~keV(FWHM) or 0.47~keV($\sigma$) in both the anode and cathode spectra.
The 1$\sigma$ resolution of the energy difference ($E_{Pt}-E_{Al}$) is 0.66~keV.
In this paper we define $\pm1.5 \sigma$ of the combined energy as the margin  in which simple averaging is utilized.

Below 23~keV, most of the events fall in this margin shown by the blue region in Figure.~\ref{avediff}.
If the energy resolution is dominated an effect of the readout electronics on each side, the energy resolution will improve by $1/\sqrt{2}$ by averaging the energies measured by the anode side and cathode side.
The improved energy resolution of 0.76~keV in the averaged spectrum is highly consistent with this number.

However, when the difference is larger than the threshold, the same method is no longer useful and only results in a degradation of energy resolution.
This is due to the tail structure shown in the red region in Figure.~\ref{avediff}.
At high energies, most events do not fall in the margin and it would degrade the lack of detection efficiency at high energy if we only use the events in the margin.
To achieve the best possible energy resolution while mitigating the loss of events, this tail structure must be corrected.

\subsection{Tail angle $\theta$ and its distribution}
\begin{figure}[htb]
\begin{center}
\includegraphics[width=\hsize]{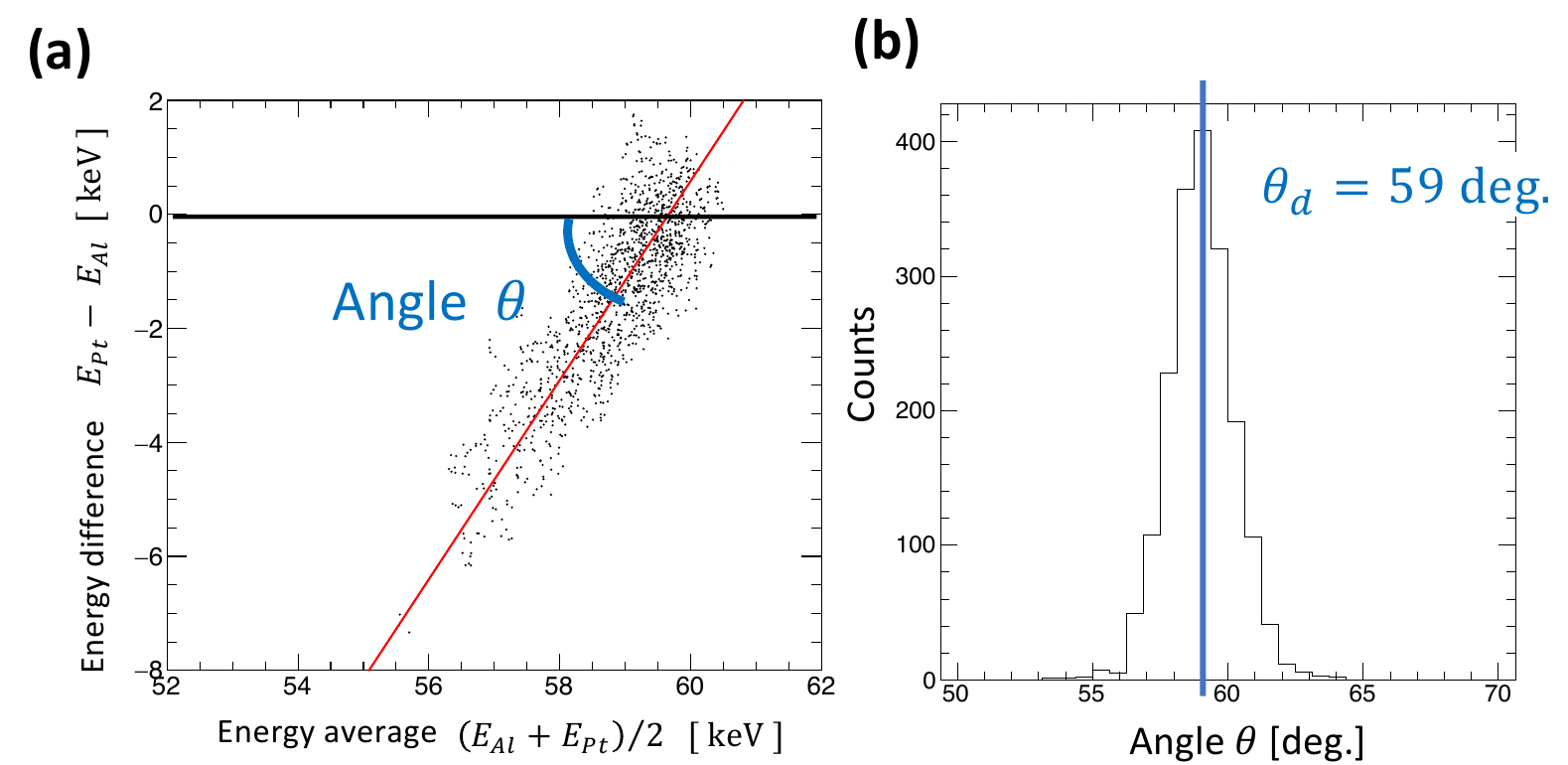}
\includegraphics[width=\hsize]{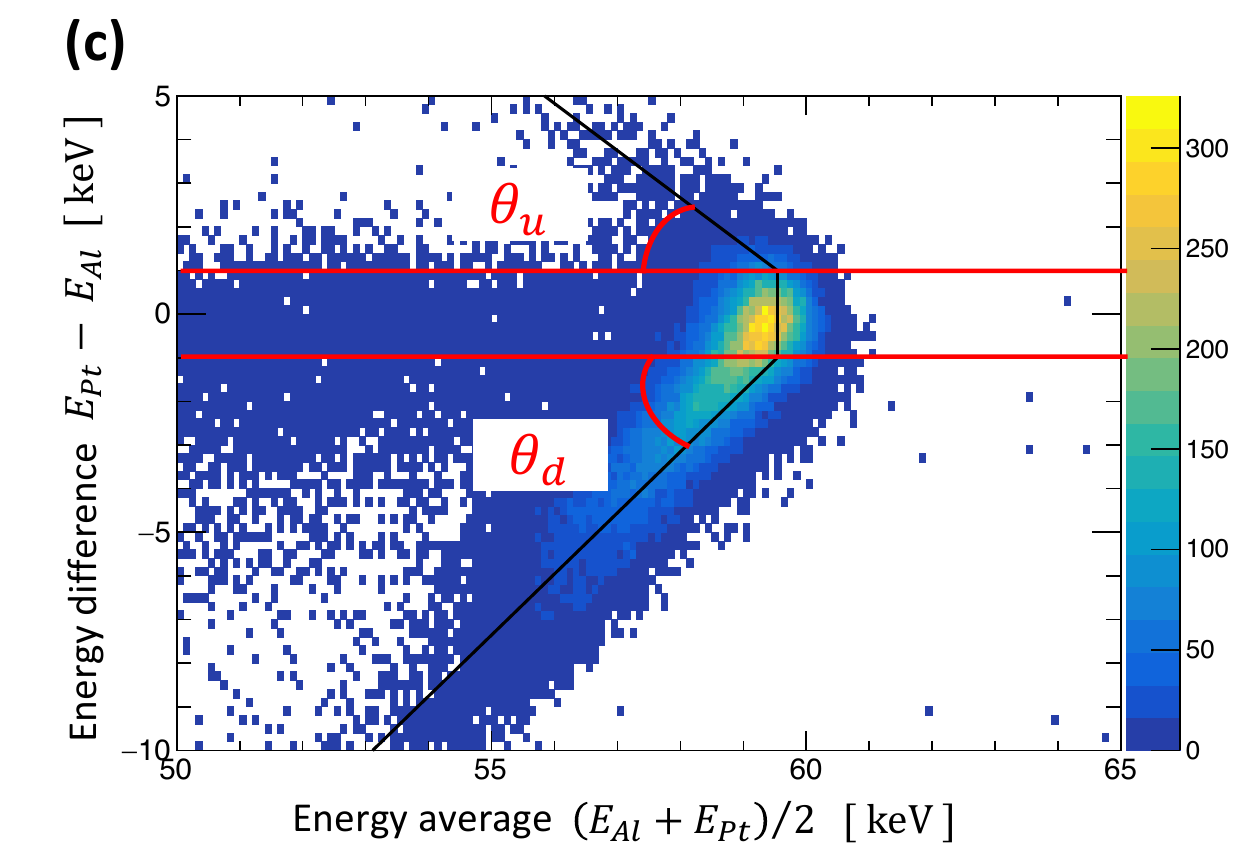}
\caption{(a)The definition and (b) distribution of tail angle $\theta$. (a) is the scatter plot of energy average and difference for a single pixel. The red line is the first principal component axis of the data points, i.e. the direction that maximizes the data variance. A tail angle $\theta$ is defined by the angle between this red line and the X-axis. The calculation is done on every pixel over the detector. (c) The reconstructed energy lines obtained from the tail angles $\theta$.}
\label{theta_ex}
\end{center}
\end{figure}

To correct the tail structure, we define a tail angle $\theta$ as the angle between the horizontal axis and the tail structure axis. 
To calculate the angle, principle component analysis (PCA) is used and the tail structure axis is defined as the first principal component axis, i.e. the direction that maximizes the data variance along the axis (Figure \ref{theta_ex}(a)).
For the actual analysis, we utilized the TPrincipal class in ROOT framework\cite{brun1997}.
To apply this framework, counts are distributed randomly in each bin of a 2-dimensional histogram of the energy average and difference of both sides and bins with less than three counts are ignored.
We calculate the tail angle for every pixel, using principal component analysis, and make a distribution of the tail angle for all pixels (Figure \ref{theta_ex}(b)).
The standard deviation is 1.2~degrees, which is approximately 2\% of the peak value.
From this result we assume theta is constant across pixels and independent of measured energy for a given branch on the plot, and is well represented by its peak value.
By using the peak value of the tail angle $\theta_{d}=59.0^{\circ}$, we can reconstruct the correct energy.

\subsection{Reconstruction algorithm and improved spectra}
The incident energy reconstruction algorithm used in this paper is as follows:\\
If the absolute energy difference between the anode and cathode energies is
\begin{itemize}
\item  within 1.0 keV:\ \ $|E_{Pt}-E_{Al}| < 1.0$ keV \\
The reconstructed energy is equal to the average energy.
\item less than $-1.0$ keV:\ \ $E_{Pt}-E_{Al} < -1.0$ keV\\
The reconstruction function is the linear combination of the difference and the average of the energies, with the reconstructed energy contour line aligned parallel to the tail-like structure. Tail angles are calculated by PCA in each pixel of the detector. The peak of the distribution for all pixels is used as the representative value $\theta_{d} = 59.0^{\circ}$.
\item greater than 1.0 keV:\ \ $E_{Pt}-E_{Al} > 1.0$ keV\\
The tail angle $\theta_{u}$ is determined as 55.3$^{\circ}$ by the PCA on the energy "average vs difference" plot summed over the detector. It is assumed that the tail structure in this case is caused by the events which photons interact very close to the cathode side\cite{salcin2014} and the percentage of this event is 5.3\%. 

\end{itemize}

\begin{figure}[htb]
\begin{center}
\includegraphics[width=\hsize]{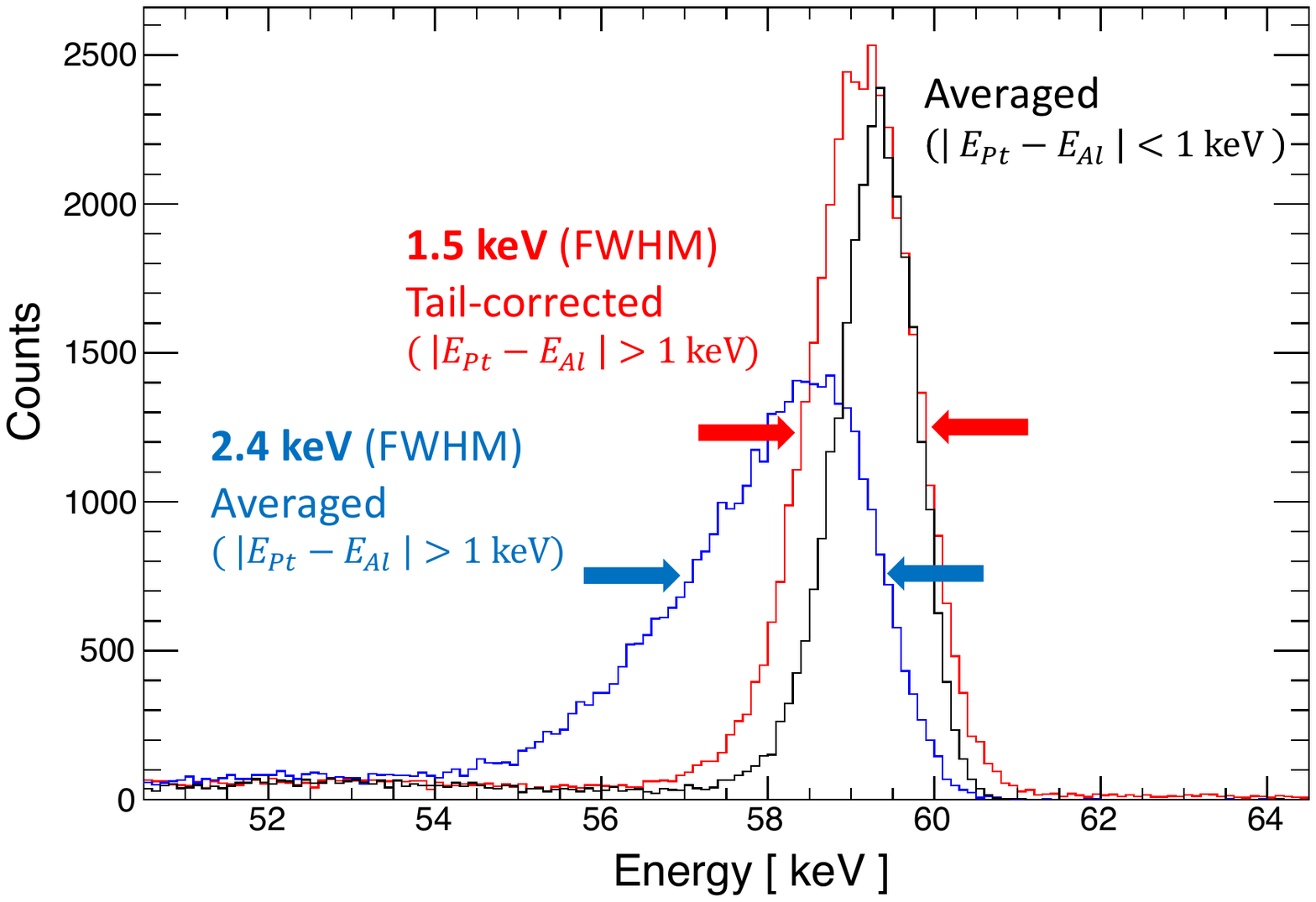}
\caption{Spectrum for the 60~keV peak (Red) processed by the tail correction method and (Blue) processed by averaging method when the energy difference is greater than 1.0~keV. (Black) Spectrum for the 60~keV peak processed by averaging method when the energy difference is within 1.0~keV(Figure~\ref{ave_when_diff_small}).
The energy resolution is improved when using the tail correction method and the lack of efficiency at high energy is mitigated.}
\label{reconst_method}
\end{center}
\end{figure}

By applying this method, the energy resolution of 60~keV peak is improved.
In Figure~\ref{reconst_method}, 60~keV peaks in the averaged spectrum (blue) and the spectrum obtained by the above method (red) with an energy difference greater than 1~keV are compared.
The averaged spectrum with an energy difference within 1~keV (black) is also superimposed.
While averaging fails to correct the tail structure, the method introduced in this section allows to mitigate the degradation of energy resolution even when the energy difference is large.
By using this tail-corrected spectrum, the lack of efficiency at high energy is mitigated while keeping the energy resolution.

\section{Properties of charge splitting}

\subsection{The Dependence on the incident photon energy}
\begin{figure}[htb]
\begin{center}
\includegraphics[width=1.0\hsize]{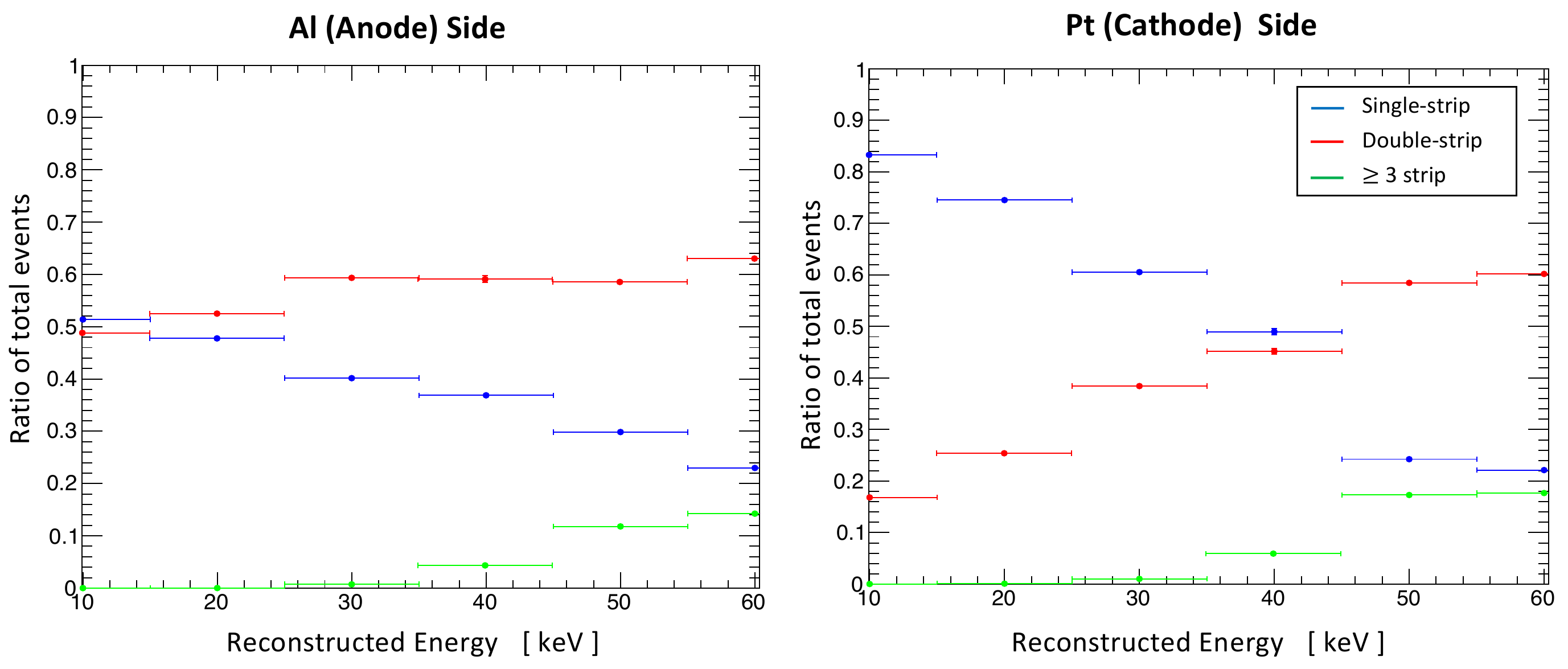}
\caption{The relationship between the reconstructed incident photon energy and the prevalence of charge sharing for ${}^{241}$Am. As the incident photon energy increases, the number of charge sharing channels increases on each side. The energy threshold of each strip is set at 1.5 keV.}
\label{rec_strip}
\end{center}
\end{figure}

The energy of incident photons can be properly reconstructed by using the method discussed in the previous section. This makes it possible to accurately assess the energy dependence of charge shared events.
Figure \ref{rec_strip} shows the incident photon energy and the degree of charge sharing between adjacent strips. As the incident photon energy increases, the charge shared event increases for both sides.
It is thought that since the initial charge cloud size is roughly proportional to the incident energy (due to electrostatic repulsion) the fraction of multiple events increases for higher energies.
There is also the effect that multiple events are more likely to detect at high energies because the energy threshold of each strip is set at 1.5 keV

\subsection{The Dependence on the depth of photon interaction}\label{dev}
\begin{figure}[htb]
\begin{center}
\includegraphics[width=1.0\hsize]{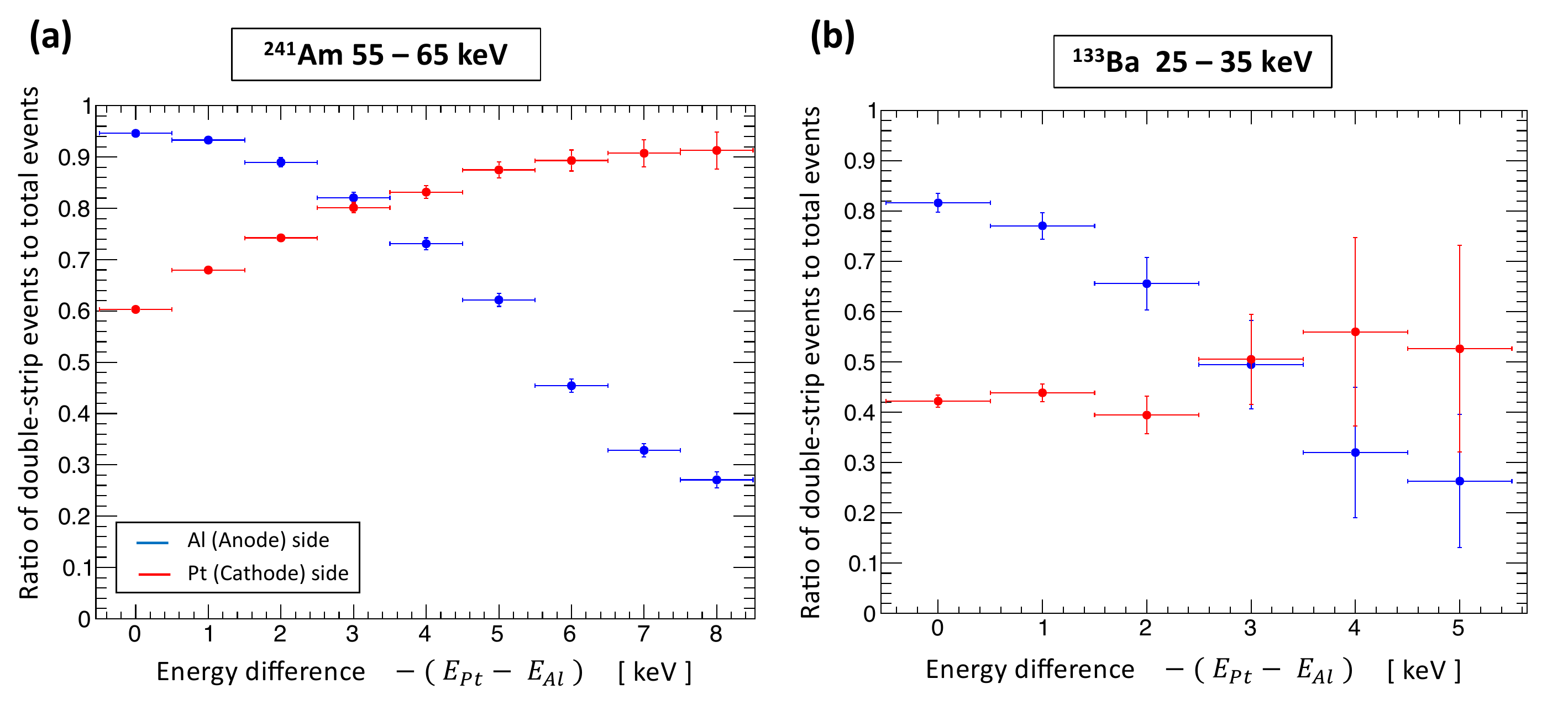}
\caption{The relationship between the depth of photon interaction and the degree of charge sharing. (a) ${}^{241}$Am 60 keV peak and (b) ${}^{133}$Ba 30 keV peak events are selected. The X-axis, energy difference between the anode and the cathode side, is bigger, this means that photons interact near the anode side. The Y-axis, ratio of double-strip event, is defined by the number of (double-strip events)/ (single and double-strip events)}
\label{depvs}
\end{center}
\end{figure}

We examine the relationship between the depth of photon interaction and the ratio of charge shared events.
It is to be noted that the difference between the cathode and anode side energy represents the depth of photon interaction.
If photons interact near the surface of the anode side, more holes are trapped and a lower energy than expected is observed in the cathode side.
Figure \ref{depvs} represents the correlation of the energy difference (i.e. photon interaction depth) and ratio of charge shared events.
As photons interact near the surface of the anode side, charge shared events at the cathode side increase.
This indicates that the longer the charge takes to travel to the electrodes, the more likely it is that a charge shared event occurs.

\section{Imaging Performance}
\subsection{Setup for evaluating imaging performance}
\begin{figure}[htb]
\begin{center}
\includegraphics[width=\hsize]{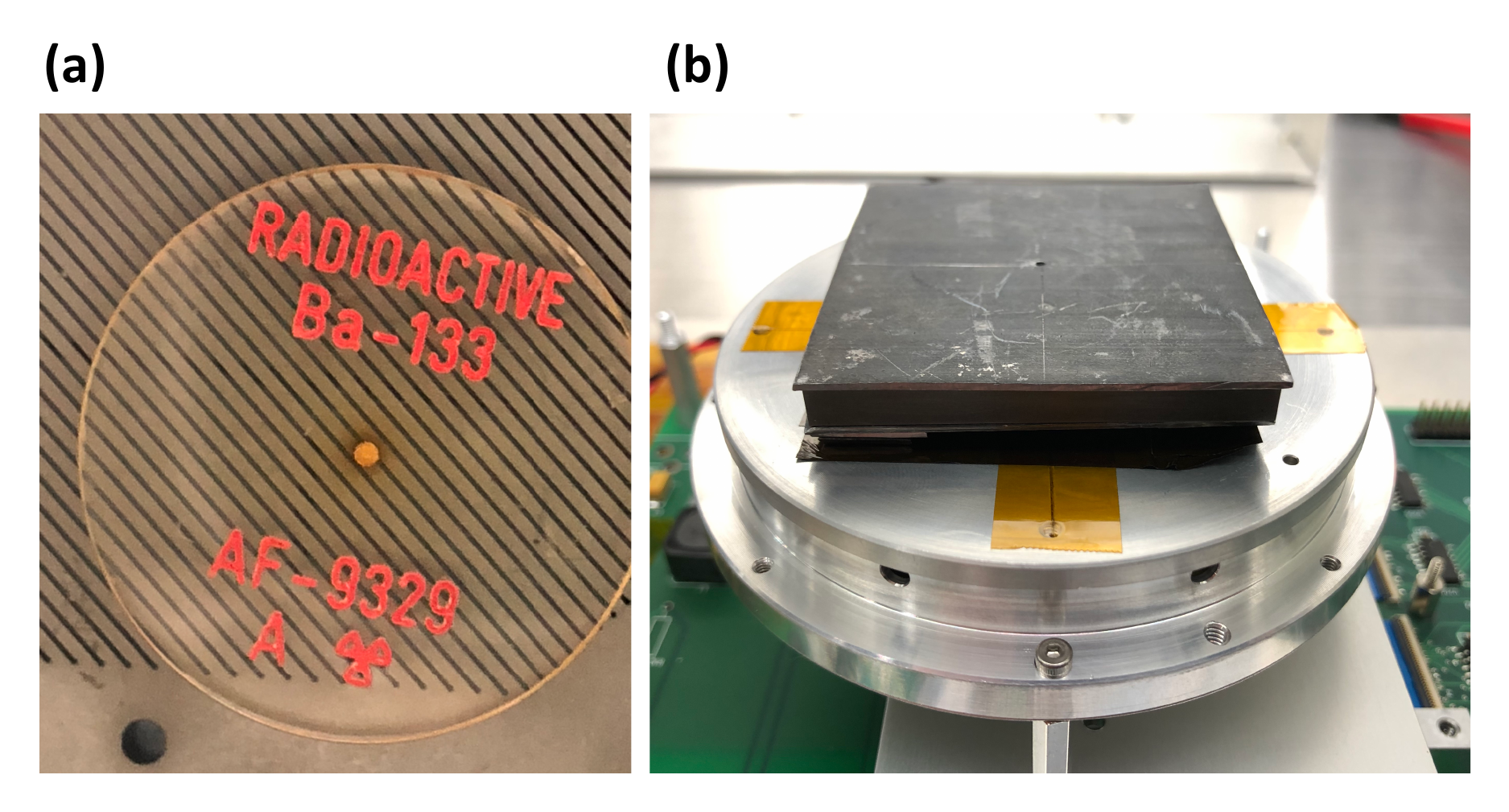}
\caption{(a) A ${}^{133}$Ba radioisotope source is covered with a 100 $\mathrm{\mu}$m slit in tungsten. (b) A graded-Z shield in lead, tin and copper is installed under the pinhole. The thickness of lead and tin is 1 mm and of copper is 0.5 mm.}
\label{img_setup_pic}
\end{center}
\end{figure}

Our previous study shows that by using information on charge sharing between adjacent channels, sub-strip resolution, finer than the strip width of 60 $\mathrm{\mu}$m, can be obtained.
To improve and confirm the sub-strip position reconstruction method, we constructed a simple imaging pinhole system.
As a target source, we used a ${}^{133}$Ba radioisotope source with a diameter of 1~mm in combination with a 100~$\mathrm{\mu}$m slit in 0.5 mm thickness tungsten (Figure \ref{img_setup_pic}(a)). 
We also developed a 100 $\mathrm{\mu}$m diameter pinhole collimator in 5 mm thickness tungsten. This is a knife-edge type of pinhole and has an opening angle of 40 degrees. Thus for high energy photons, the effective pinhole diameter is a little bigger, 120 $\mathrm{\mu}$m for 30 keV (90\% absorption) for example.

\begin{figure}[htb]
\begin{center}
\includegraphics[width=\hsize]{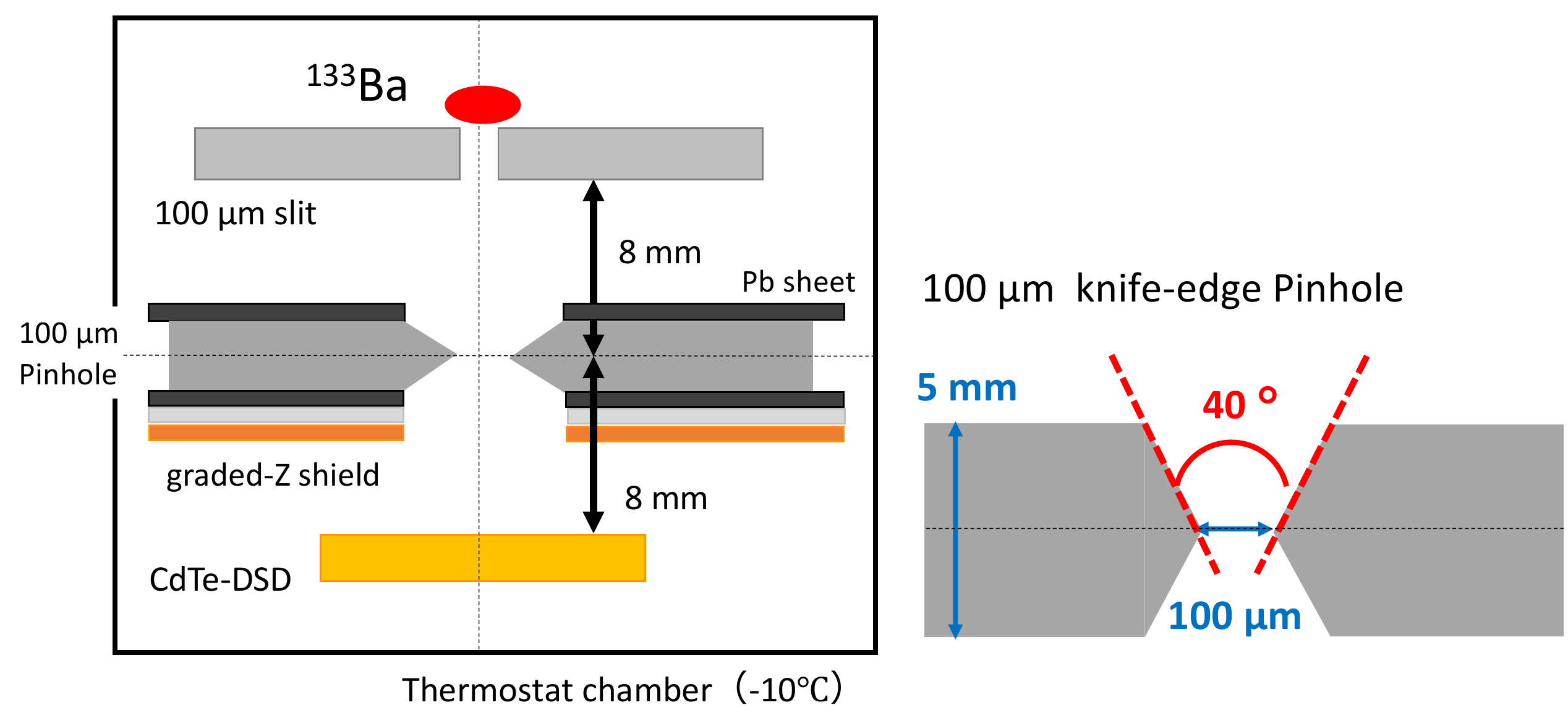}
\caption{Setup configuration of the fine pinhole imaging system. The detector, the radioactive isotope, and the pinhole collimator are in a thermostat chamber and kept at $-$10 degrees Celsius. From outside the chamber, power and bias voltage are supplied.}
\label{img_setup}
\end{center}
\end{figure}

The experimental setup is depicted in Figure \ref{img_setup}.
Since the distance from the target source to the pinhole collimator and from the pinhole collimator to the detector is the same, the image is approximately the same size as the source.
To suppress background, we installed a graded-Z shield under the pinhole made out of lead, tin and copper  (Figure \ref{img_setup_pic}(b)).

\begin{figure}[htb]
\begin{center}
\includegraphics[width=\hsize]{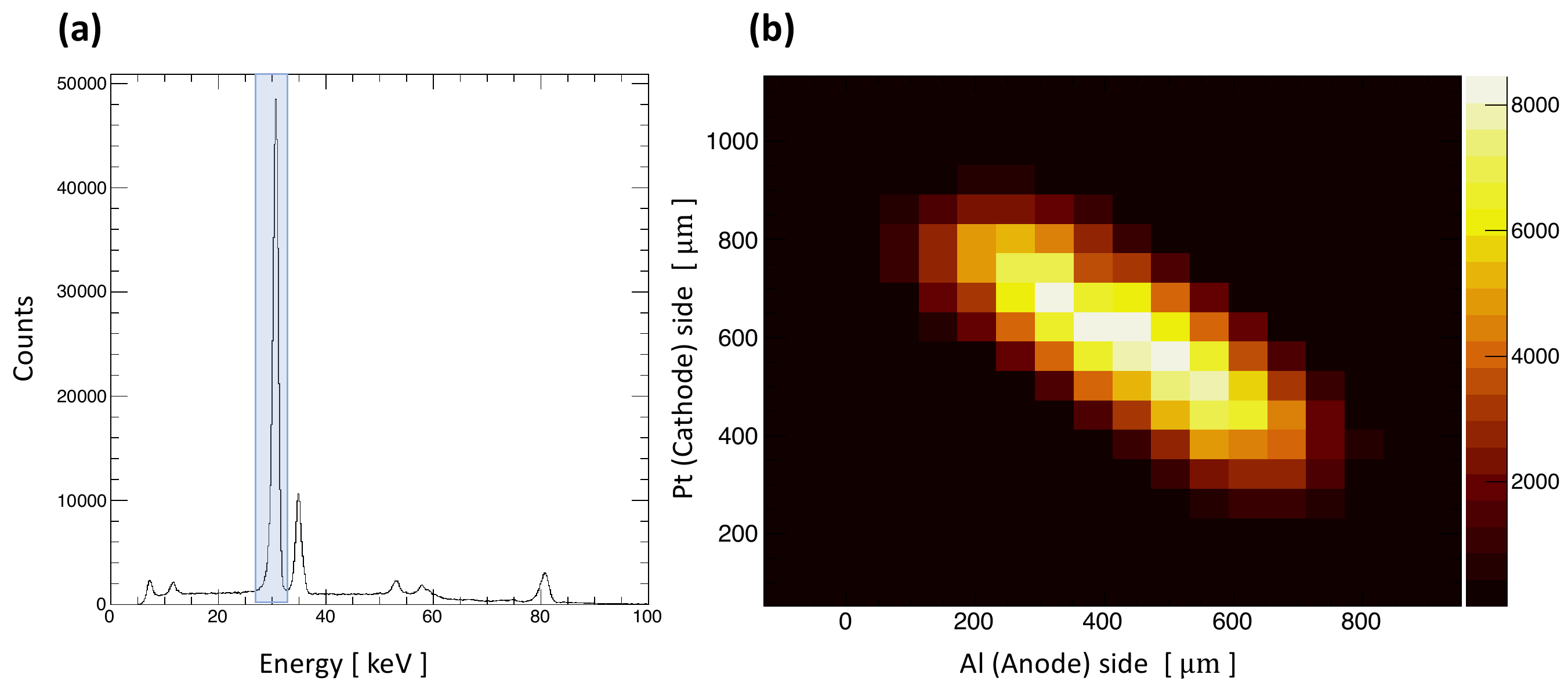}
\caption{(a) ${}^{133}$Ba spectra of reconstructed energy. (b) 60 $\mathrm{\mu}$m strip resolution image using 28-32 keV events(blue region of (a)).}
\label{img_ini}
\end{center}
\end{figure}

The energy spectrum and strip-resolution image are presented in Figure \ref{img_ini}.
The image is reconstructed by using 28-32 keV events and we only use single and double strip charge shared events. Here, the strip where the signal is detected is assigned as the photon interaction position for each one strip event. For double strip events, the strip where the bigger signal is detected is simply used.
However, 60 $\mathrm{\mu}$m strip resolution is insufficient to image the 100 $\mathrm{\mu}$m target source in detail.
So, to realize sub-strip resolution and improve this image, charge sharing information between adjacent channels should be taken into account.
While we have already developed a method to realize sub-strip resolution\cite{Furukawa2018}, we now developed a new reconstruction method to improve imaging performance by taking into account charge cloud spreading as discussed in section \ref{dev}.

\subsection{Improved position reconstruction method to realize sub-strip resolution}
\begin{figure}[htb]
\begin{center}
\includegraphics[width=\hsize]{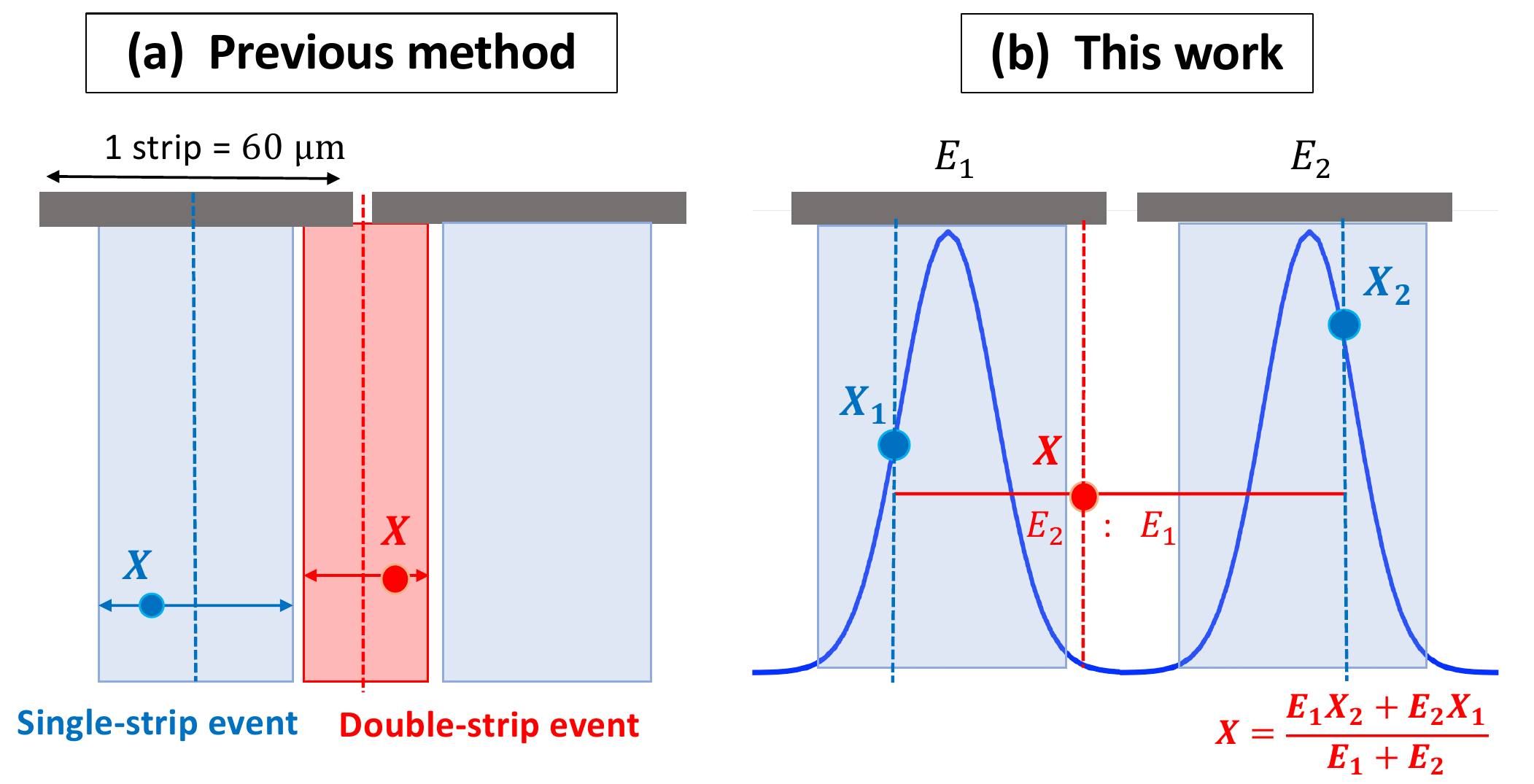}
\caption{Position reconstruction methods. (a) Previous method: position is randomly assigned according to a uniform probability distribution within each defined region. (b) This work: position is randomly assigned according to a gaussian probability distribution for each strip.
}
\label{reconst_imgmethod}
\end{center}
\end{figure}

The previous and improved method are shown in Figure \ref{reconst_imgmethod}.
The previous method\cite{Furukawa2018} divides a region into single- and double-strip charge shared event region on the assumption that events are split only in the double-strip event region around the electrode gap.
The incident position of the photon is randomly assigned according to a uniform probability distribution in each region.
The width of the single- and double-strip regions is determined according to the number of double-strip events divided by the sum of the numbers of single- and double-strip events.

The improved method is that the incident position of the photon is randomly assigned according to a gaussian probability distribution at each strip.
From the charge spread property discussed in section \ref{dev}, we assume that the charge clouds created by incident photon diffuse following the Fick equation\cite{engel2012} and choose a gaussian model.
The mean of the gaussian is defined as the center of the strip and the sigma is defined by calculating the ratio of charge shared events as in the previous method.
For the anode side, 28-32 keV energy events, the number of single-strip events is 62.5\%.
So two sigmas correspond to 60 $\mathrm{\mu m} \times $ 62.5\% = 37.5 $\mathrm{\mu}$m.
Conversely for the cathode side, two sigmas correspond to 60 $\mathrm{\mu m} \times $ 40.8\% = 24.5 $\mathrm{\mu}$m.
In the case of a double-strip charge shared event, the position interaction is reconstructed in two steps.
First, as for single-strip events, each strip position is randomly assigned according to a gaussian probability distribution.
Then, the position is determined by calculating the energy-weighted center of each strip position. 

\begin{figure}[htb]
\begin{center}
\includegraphics[width=\hsize]{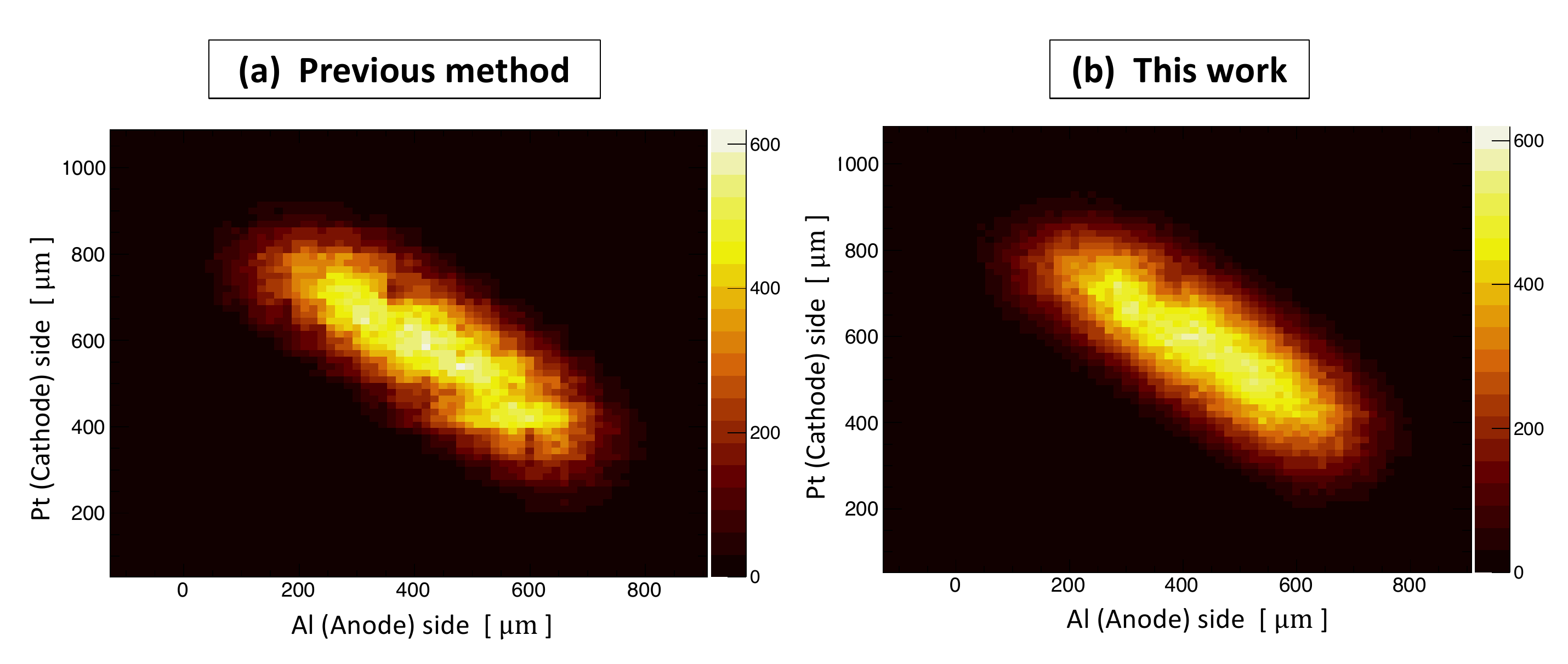}
\caption{Sub-strip reconstruction image applying the  (a) previous\cite{Furukawa2018} and (b) improved method. The artificial pixellated structure of (a) has disappeared and the 100 $\mathrm{\mu}$m slit image is reconstructed accurately.
}
\label{img_result}
\end{center}
\end{figure}

Figure \ref{img_result} shows a comparison of images obtained by the two methods of position reconstruction.
With the new, improved method, a much smoother image of 100~$\mathrm{\mu}$m slit was obtained. 
This suggests that, by using the new model of charge sharing and event reconstruction, the sub-strip resolution (smaller that the strip pitch of 60 $\mathrm{\mu}$m) can be better demonstrated. 

%

\section{Conclusion}
We have developed a 60 $\mu$m fine-pitch CdTe-DSD for the FOXSI-3 Sounding Rocket Experiment. 
In terms of spectroscopic performance, the cathode side spectrum has a tail structure at high energy peaks because of charge trapping of holes.
By applying an depth of photon interaction correction, if the energy difference between the anode and cathode is within 1 keV, energy resolutions (FWHM) of 0.76 keV and 1.0 keV at 14 keV and 60 keV can be obtained, respectively. For events in which the energy difference is greater than 1 keV, the energy resolution at the 60 keV peak is improved to 1.5 keV compared to 2.4 keV obtained by averaging both sides of the energy.
This correction also makes it possible to accurately assess the energy dependence of charge shared events.
The incident photons of high energy or interact far from the electrode are more likely to split charge at the electrodes, giving rise to an energy dependence and spreading property of charge cloud.
By taking into account this knowledge and information about charge sharing between adjacent channels, we developed a new image reconstruction method.
Using this method, a 100 $\mu$m slit image was reconstructed and sub-strip resolution finer than the strip width of 60 $\mu$m is confirmed.

\section*{Acknowledgement}
The authors would like to thank Pietro Caradonna for his critical reading of the paper. This work was supported by JSPS KAKENHI Grant Number 24244021, 16H03966, 16H02170, 18H05463 and 20H00153. This work was also supported by FoPM, WINGS Program, the University of Tokyo.


\bibliography{main}

\end{document}